\documentstyle [12pt,epsf] {article}

\def\be{\begin{equation}}
\def\ee{\end{equation}}
\def\ba{\begin{array}}
\def\ea{\end{array}}
\def\bea{\begin{eqnarray}}
\def\eea{\end{eqnarray}}
\def\GeV{{\rm GeV}}
\def\tr{{\rm tr}}
\def\thefootnote{\fnsymbol{footnote}}
\def\chib{{\bar\chi}}
\def\psib{{\bar\psi}}
\def\nn{\nonumber}

\def\wS{S}
\def\wT{T}
\def\sS{{\cal S}}
\def\sT{{\cal T}}

\def\NPB#1#2#3{{Nucl.~Phys.} {\bf{B#1}} (19#2) #3}
\def\PLB#1#2#3{{Phys.~Lett.} {\bf{B#1}} (19#2) #3}
\def\PRD#1#2#3{{Phys.~Rev.} {\bf{D#1}} (19#2) #3}

\if@twoside\oddsidemargin25mm
\else\oddsidemargin25mm\evensidemargin25mm\marginparwidth25mm\fi%
\begin{document}
\begin{titlepage}
{\sf
\begin{flushright}
{TUM--HEP--282/97}\\
{SFB--375/201}\\
{July 1997}
\end{flushright}}
\vfill
\vspace{-1cm}
\begin{center}
{\large \bf Supersymmetry Breaking and Soft Terms in M-Theory}

\vskip 1.2cm
{\sc H. P. Nilles$^{1,2}$, {\ }M. Olechowski$^{1,\ast}$ {\ and} 
{\ \sc M. Yamaguchi$^{1,\dagger}$}\\
\vskip 1.5cm
{\em $^1$Institut f\"{u}r Theoretische Physik} \\
{\em Physik Department} \\
{\em Technische Universit\"at M\"unchen} \\
{\em D--85747 Garching, FRG}
\vskip 1cm
{\em $^2$Max--Planck--Institut f\"ur Physik} \\
{\em ---Werner--Heisenberg--Institut---}\\
{\em D--80805 M\"{u}nchen, FRG}}
\end{center}
\vfill

\thispagestyle{empty}

\begin{abstract}

We investigate gaugino condensation in the framework of the strongly 
coupled heterotic $E_8 \times E_8$ string (M--theory). Supersymmetry 
is broken in a hidden sector and gravitational interactions induce soft 
breaking parameters in the observable sector.
The resulting soft masses are of order of the gravitino mass. 
The situation is similar to that in the weakly coupled 
$E_8 \times E_8$ theory with one important difference: 
we avoid the problem of small gaugino masses which are now 
comparable to the gravitino mass.

\end{abstract}

\vskip 5mm \vskip0.5cm
\hrule width 5.cm \vskip 1.mm
\noindent
{\small\small ${}^{\ast}$ On leave of absence from 
Institute of Theoretical Physics, Warsaw University, Poland.\\}
{\small\small ${}^{\dagger}$ On leave of absence from 
Department of Physics, Tohoku University, Sendai 980--77, Japan.}
\end{titlepage}

\def\thefootnote{\arabic{footnote}}
\setcounter{footnote}{0}

Phenomenological applications of string theory have mainly concentrated on 
the weakly coupled heterotic $E_8 \times E_8$ string. Models with realistic
gauge groups and particle content have been constructed. Qualitatively 
the unification of gauge and gravitational couplings can be understood 
\cite{GHMR} 
with some moderate uncertainties in the relation between the string 
scale $M_{\rm string}$ and the grand unified scale $M_{\rm GUT}$. 
Supersymmetry is broken dynamically via gaugino condensation 
\cite{DIN,DRSW} 
leading to a hidden sector supergravity model at low energies with specific 
predictions for the soft breaking parameters including problematically 
small gaugino masses
\cite{BIM}. 
At the moment we do not possess a convincing mechanism to resolve this 
problem.

Recently there have been attempts to study string theories in the region 
of intermediate and strong coupling. The strongly coupled version of the 
$E_8\times E_8$ theory is believed to be an orbifold of 11--dimensional 
M--theory, an interval in $d=11$ with $E_8\times E_8$ gauge fields 
restricted to the two $d=10$ dimensional boundaries respectively 
\cite{HW}. 
Applied to the question of unification 
\cite{W} 
the following picture emerges: the GUT--scale 
$M_{\rm GUT} = 3\times 10^{16}$~GeV is identified with 
$1/R$ where $V=R^6$ is the volume of compactified six--dimensional space. 
$\alpha_{\rm GUT} = 1/25$ and the correct value of the $d=4$ reduced 
Planck mass $M_P = 2.4 \times 10^{18}$ GeV can be obtained by choosing 
the length of the $d=11$ interval to be $R_{11}\approx 6R$. The fundamental 
mass scale of the $d=11$ theory $M_{11}= \kappa^{-2/9}$ (with $\kappa$ 
the $d=11$ Einstein gravitational coupling) has to be chosen a factor 
2 larger than $M_{\rm GUT}$ and at that scale 
$\alpha_{\rm string}=g^2_{\rm string}/4\pi$  is of order unity. This then 
represents a rather natural framework for the unification of coupling 
constants. The large compactification radius can be used to solve the 
strong CP problem  
\cite{BD,C}.

As in the weakly coupled theory, supersymmetry might be broken dynamically 
by gaugino condensation in the hidden $E_8$ on one boundary of space--time 
\cite{H}. 
Gravitational interactions will play the role of messengers to the 
observable sector at the opposite boundary.

In this paper we shall discuss this mechanism in detail and compute the 
predictions for the low energy effective theory. We find results very 
similar to the situation in the $d=10$ weakly coupled case, with one 
notable exception: {\it gaugino masses are of comparable size to the 
gravitino mass, thus solving the problem of small gaugino masses that 
occurred in the weakly coupled case}.

Let us first review gaugino condensation in the $d=10$ weakly coupled 
$E_8 \times E_8$ theory. We shall start from the $d=10$ effective field 
theory and go to $d=4$ dimensions via the method of reduction and truncation 
explained in ref.\ \cite{W2}. In string theory compactified on an 
orbifold this would describe the dynamics of the untwisted sector. 
We retain the usual moduli fields $\wS$ and $\wT$ as well as matter 
fields $C_i$ that transform nontrivially under the observable sector 
gauge group. In this approximation, the K\"ahler potential is given by 
\cite{W2,DIN2}
\be
G = - \log (\wS + \bar \wS) - 3 \log (\wT+\bar \wT -2C_i\bar C_i) + 
\log \left| W \right|^2  
\label{eq:G}  
\ee
with superpotential
\be
 W(C) = d_{ijk} C_iC_jC_k   
\ee
and the gauge kinetic function is given by the dilaton field
\be
f = \wS \,. 
\ee
We assume in the hidden sector the appearance of a gaugino condensate
\be 
\left<\chi\chi\right>  = \Lambda^3 
\ee
where $\Lambda$ is the renormalization group invariant scale of the 
confining hidden sector gauge group. The gaugino condensate appears 
in the expression for the auxiliary components of the chiral superfields 
\cite{FGN} 
\be
F_j = \left( G^{-1} \right)_j^k 
      \left( \exp (G/2) G_k +{1\over 4} f_k (\chi\chi) \right) + \ldots 
\ee
which are order parameters for supersymmetry breakdown. Minimizing the 
scalar potential we find $F_\wS = 0$, $F_\wT \ne 0$ and a vanishing 
cosmological constant. Supersymmetry is broken and the gravitino mass 
is given by
\be
m_{3/2} = \frac{\left< F_T \right>}{\wT + {\bar \wT}} 
\approx {\Lambda^3 \over M_P^2} 
\label{eq:m32}
\ee
and $\Lambda = 10^{13}$ GeV would lead to a gravitino mass in the 
TeV -- range. A first inspection of the soft breaking terms in the 
observable sector gives a disturbing result. They vanish in this 
approximation. Scalar masses are zero because of the no--scale structure 
in (\ref{eq:G}) (coming from the fact that we have only included fields 
of modular weight $-1$  under T--duality in this case) 
\cite{BIM}. 
In a more general situation we would get scalar masses $m_0$ comparable to 
the gravitino mass $m_{\rm 3/2}$ and the above result  $m_0=0$ is just an 
artifact of the chosen approximation at the classical level. Gaugino 
masses $m_{1/2}$ are given by
\be
m_{1/2} = \frac{ {\partial f \over \partial \wS} F_\wS 
                +{\partial f \over \partial \wT} F_\wT }
               {2 {\rm Re} f}                          
\label{eq:m12}
\ee
and with $f = \wS$ and $F_\wS = 0$ we obtain $m_{1/2} = 0$. One loop 
corrections will change this picture as can be seen already by an 
inspection of the Green--Schwarz anomaly cancellation counter terms, 
as they modify $f$ at one loop. In the simple example of the so--called
standard embedding with gauge group $E_6 \times E_8$ we obtain 
\cite{DIN2,IN}
\be
f_6 = \wS +\epsilon \wT\,;
\qquad\qquad 
f_8 = \wS - \epsilon \wT\,.
\label{eq:f6f8}
\ee
This dependence of $f$  on $\wT$ will via (\ref{eq:m12}) lead to 
nonvanishing gaugino masses which, however, will be small compared 
to $m_{3/2}$ and $m_0$ since $\epsilon \wT$ is considered 
a small correction to the classical result. This might be   
problematic when applied to the supersymmetric extension of 
the standard model.

Let us now move to the strongly coupled $E_8 \times E_8$ -- $M$--theory. 
The effective action is given by 
\cite{HW}
\bea
L 
\!\!&=&\!\!
{1\over \kappa^2} \int
d^{11}x \sqrt{g}
\left[
       - \frac{1}{2}R 
       - \frac{1}{2} \psib_I \Gamma^{IJK} D_J
            \left( \frac{\Omega + {\hat\Omega}}{2} \right) \psi_K 
       - \frac{1}{48} G_{IJKL} G^{IJKL}
\right.
\nn\\
&&\qquad\qquad\qquad
       - \frac{\sqrt{2}}{384} 
            \left( \psib_I \Gamma^{IJKLMN} \psi_N
                  +12 \psib^J \Gamma^{KL} \psi^M \right)
            \left( G_{JKLM} + {\hat G}_{JKLM} \right)
\nn\\ 
&&\qquad\qquad\qquad
       - \left. \frac{\sqrt{2}}{3456}
            \epsilon^{I_1 I_2 \ldots I_{11}} C_{I_1 I_2 I_3}
            G_{I_4 \ldots I_7} G_{I_8 \ldots I_{11}}
\right]
\\
&&
+ \frac{1}{2\pi(4\pi\kappa^2)^{2/3}}  \int 
d^{10}x \sqrt{g}
\left[ 
    - \frac{1}{4} F^a_{AB} F^{aAB} 
       - \frac{1}{2} \chib^a \Gamma^AD_A ({\hat\Omega}) \chi^a
\right.
\nn\\
&&\qquad\qquad\qquad\qquad\qquad\quad
\left.
       - \frac{1}{8} \psib_A \Gamma^{BC} \Gamma^A
            \left( F^a_{BC} + {\hat F}^a_{BC} \right) \chi^a
       + \frac{\sqrt{2}}{48}
            \left( \chib^a \Gamma^{ABC} \chi^a\right){\hat G}_{ABC11}
\right].
\nn
\eea
Compactifying to $d=4$ we obtain 
\cite{W}
\be
G_N =   8\pi \kappa_4^2 = {\kappa^2 \over 16\pi^2 V \rho}\,,
\qquad\qquad
\alpha_{GUT} = {(4\pi\kappa^2)^{2/3} \over 2V}
\label{eq:GN}
\ee
with $V= R^6$ and $\pi\rho= R_{11}$. Fitting $G_N$ and $\alpha_{GUT}=1/25$ 
then gives $R_{11} M_{11}\approx 12$ and 
$M_{11} R\approx 2$ ($M_{11} \approx 6\times 10^{16}$ GeV).

The rather large value of the $d=4$ reduced Planck Mass  
$M_{P}= \kappa_4^{-1}$ is obtained as a result of the fact that $R_{11}$
is large compared to $R$.

We now perform a compactification using the method of reduction and 
truncation as above. For the metric we write
\be
g_{MN} = 
\left(
\ba{ccc}
e^{-\gamma} e^{-2\sigma} g_{\mu\nu} & & \\
 & e^\sigma \delta_{mn} & \\
 & & e^{2\gamma} e^{-2\sigma} 
\ea
\right)
\ee
with 
$M,N = 1 \ldots 11$; $\mu,\nu = 1 \ldots 4$; $m,n = 5 \ldots 10$; 
$2R_{11} = 2\pi\rho  = M_{11}^{-1} e^\gamma e^{-\sigma}$  and
$V= e^{3\sigma} M_{11}^6$.  At the classical level this leads to a 
K\"ahler potential as in (\ref{eq:G}) 
\be
K = - \log (\sS + {\bar \sS}) - 3 \log (\sT+{\bar \sT} -2C_i {\bar C_i})
\ee
with
\bea
\sS &=& \frac{2}{\left( 4\pi \right)^{2/3}} 
\left( e^{3\sigma} \pm i 24 \sqrt{2} D \right)
\,,
\\
\sT &=& \frac{\pi^2}{\left( 4\pi \right)^{4/3}}
\left( e^{\gamma} \pm i 6 \sqrt{2} C_{11} \right)
\eea
where $D$ and $C_{11}$ fields are defined by
\bea
\frac{1}{4!} e^{6\sigma} G_{11\lambda\mu\nu} &=& 
\epsilon_{\lambda\mu\nu\rho}\left(\partial^\rho D \right)
\,,
\\
C_{11 i {\bar j}} &=& C_{11} \delta_{i \bar j}
\eea
and $x^i$ ($x^{\bar j}$) is the holomorphic (antiholomorphic) coordinate 
of the Calabi--Yau manifold. The imaginary part of $\sS$ (Im$\sS$) 
corresponds to the model independent axion, and the gauge kinetic 
function is $f = \sS$. This is very similar to the weakly coupled case. 
Before drawing any conclusion from these formulae, however, we have to 
discuss a possible obstruction at the one loop level. It can be understood 
from the mechanism of anomaly cancellation 
\cite{W}. 
For the 3--index tensor field $H$ in $d=10$ supergravity to be well defined 
one has to satisfy $dH = \tr F_1^2 + \tr F_2^2 - \tr R^2 = 0$ 
cohomologically. In the simplest case of the standard embedding one 
assumes $\tr F_1^2 = \tr R^2$ locally and the gauge group is broken to 
$E_6 \times E_8$. Since in the M--theory case the two different gauge 
groups live on the two different boundaries of space--time such a 
cancellation point by point is no longer possible. We expect nontrivial 
vacuum expectation values (vevs) of 
\be
(dG) \propto \sum_i \delta(x^{11} - x^{11}_i) 
\left( \tr F_i^2 - {1\over 2} \tr R^2 \right)
\ee
at least on one boundary ($x^{11}_i$ is the position of $i$--th boundary). 
In the case of the standard embedding we would have 
$\tr F_1^2 - {1\over 2} \tr R^2 = {1\over 2} \tr R^2$ on one and 
$\tr F^2_2 - {1\over 2} \tr R^2 = - {1\over 2} \tr R^2$ on the other 
boundary. This might pose a severe problem since a nontrivial vev  of 
$G$ might be in conflict with supersymmetry ($G_{11ABC}=H_{ABC}$). 
The supersymmetry 
transformation law in $d=11$ reads  
\be
\delta \psi_M 
=
D_M\eta + \frac{\sqrt{2}}{288} G_{IJKL} 
          \left( \Gamma_M^{IJKL} - 8 \delta_M^I \Gamma^{JKL} \right) \eta
+ \ldots
\label{eq:dpsiM}
\ee
Supersymmetry will be broken unless e.g.\ the derivative term $D_M\eta$ 
compensates the nontrivial vev of $G$. Witten has shown 
\cite{W} 
that such a cancellation can occur and constructed the solution in the 
linearized approximation (linear in the expansion parameter $\kappa^{2/3}$) 
which corresponds to the large $T$--limit in the weakly coupled 
theory\footnote
{For a discussion beyond this approximation in the weakly coupled 
case see ref.\ 
\cite{NS}.
}. 
The supersymmetric solution leads to a nontrivial dependence of the 
$\sigma$ and $\gamma$ fields with respect to $x^{11}$: 
\be
{{\partial\gamma}\over{\partial x^{11}}} 
=
- {{\partial\sigma}\over{\partial x^{11}}} 
=
\frac{\sqrt{2}}{24}
\frac
{\int d^6x \sqrt{g} \omega^{AB}\omega^{CD}G_{ABCD}}
{\int d^6x \sqrt{g}}
\ee
where the integrals are over the Calabi--Yau manifold and $\omega$ is 
the corresponding K\"ahler form. A definition of our $\sS$ and $\sT$ 
fields in the four--dimensional theory would then require an average 
over the 11--dimensional interval. We would therefore write
\bea
\sS &=& \frac{2}{\left( 4\pi \right)^{2/3}} 
\left( e^{3\bar\sigma} \pm i 24 \sqrt{2} \bar D \right)
\,,
\\
\sT &=& \frac{\pi^2}{\left( 4\pi \right)^{4/3}}
\left( e^{\bar\gamma} \pm i 6 \sqrt{2} \bar C_{11} \right)
\label{eq:sT}
\eea
where bars denote averaging over the 11th dimension. It might be of some 
interest to note that the combination $\sS\sT^3$ is independent of 
$x^{11}$ even before this averaging procedure took place.

$\exp(3\sigma)$ represents the  volume of the six--dimensional compact 
space in units of $M_{11}^{-6}$. The $x^{11}$ dependence of $\sigma$ 
then leads to the geometrical picture that the volume of this space 
varies with $x^{11}$ and differs at the two boundaries. In the given 
approximation, this variation is linear, and for growing $R_{11}$ the 
volume on the $E_8$ side becomes smaller and smaller. At a critical 
value of $R_{11}$ the volume will thus vanish and this will provide 
us with an upper limit on $R_{11}$. For the phenomenological 
applications we then have to check whether our preferred choice of $R_{11}$
that fits the correct value of the $d=4$ Planck 
mass\footnote
{With $V$ depending on $x^{11}$ we have to specify which values 
should be used in eq.\ (\ref{eq:GN}). The appropriate choice 
in the expression for $G_N$ is the average value of $V$ 
while in the expression for $\alpha_{GUT}$
we have to use $V$ evaluated at the $E_6$ border.} 
satisfies this bound. Although the coefficients are model dependent we 
find in general that the bound can be satisfied, but that $R_{11}$ is 
quite close to its critical value. A choice of $R_{11}$ much larger than 
$(5\times 10^{15} \GeV)^{-1}$ is therefore not permitted.

This variation of the volume is the analogue of the one loop correction 
of the gauge kinetic function (\ref{eq:f6f8}) in the weakly coupled case 
and has the same origin, namely a Green--Schwarz anomaly cancellation 
counterterm. In fact, also in the strongly coupled case we find corrections 
for the gauge coupling constants at the $E_6$ and $E_8$ side.

Gauge couplings will no longer be given by the (averaged) $\sS$--field, 
but by that combination of the (averaged) $\sS$ and $\sT$ fields which 
corresponds to the $\sS$--field before averaging at the given boundary: 
\be  
f_{6,8} = \sS \pm \alpha \sT
\ee
at the $E_6$ ($E_8$) side 
respectively\footnote
{With the normalization of the $\sT$ field as in (\ref{eq:sT}), 
$\alpha$ is a quantity of order 1.}.
The critical value of $R_{11}$ will correspond to infinitely strong 
coupling at the $E_8$ side $\sS - \alpha \sT = 0$ (Notice the similarity 
to (\ref{eq:f6f8}) in the weakly coupled limit). Since we are here close 
to criticality a correct phenomenological  fit of 
$\alpha_{\rm GUT} = 1/25$ should include this correction 
$\alpha_{\rm GUT}^{-1} = \sS + \alpha \sT$ where $\sS$ and 
$\alpha \sT$ give comparable contributions. This is a difference to the 
weakly coupled case, where in $f= \wS + \epsilon \wT$ the latter 
contribution was small compared to $\wS$. Observe that this picture of 
a loop correction $\alpha \sT$ to be comparable to the tree level result 
still makes sense in the perturbative expansion, since $f$ does not 
receive further perturbative corrections beyond one loop 
\cite{SV,N}.

In a next step we are now ready to discuss the dynamical breakdown 
of supersymmetry via gaugino condensation in the strongly coupled 
M--theory picture. In analogy to the previous discussion we start 
investigating supersymmetry transformation laws in the higher--dimensional 
(now $d=11$) field theory 
\cite{H}:
\bea
\delta \psi_A
&=&
D_A\eta 
+ \frac{\sqrt{2}}{288} G_{IJKL} 
  \left( \Gamma_A^{IJKL} - 8 \delta_A^I \Gamma^{JKL} \right) \eta
\nn\\
&&- 
\frac{1}{576\pi} \left( \frac{\kappa}{4\pi} \right)^{2/3} \delta(x^{11})
  \left( \chib^a \Gamma_{BCD} \chi^a \right) 
  \left( \Gamma_A^{BCD} - 6 \delta_A^B \Gamma^{CD} \right) \eta
  + \ldots
\\
\delta \psi_{11}
&=&
D_{11} \eta + \frac{\sqrt{2}}{288} G_{IJKL}
\left( \Gamma_{11}^{IJKL} - 8 \delta_{11}^I \Gamma^{JKL} \right) \eta
\nn\\
&&+
  \frac{1}{576\pi} \left( \frac{\kappa}{4\pi} \right)^{2/3}
  \delta(x^{11}) \left( \chib^a \Gamma_{ABC} \chi^a \right) \Gamma^{ABC} \eta
  + \ldots
\eea
where gaugino bilinears appear in the right hand side of both expressions. 
It can therefore be expected that gaugino condensation breaks supersymmetry. 
Still the details have to be worked out. In  the $d=10$ example, the 
gaugino condensate and the three--index tensor field $H$ contributed to 
the scalar potential in a full square. This lead to a vanishing 
cosmological constant as well as the fact that $F_\wS =0$ at the classical 
level. Ho\v{r}ava has observed 
\cite{H} 
that a similar mechanism might be in operation in the $d=11$ theory
$$
- \frac{1}{12\kappa^2} \int_{M^{11}} d^{11}x \sqrt{g} G^2_{ABC11}
+ \frac{\sqrt{2}}{24(4\pi)^{5/3}\kappa^{4/3}}
  \int_{M^{10}} d^{10}x \sqrt{g} G_{11ABC}
           \left( \chib^a \Gamma^{ABC} \chi^a \right)
$$
\be
- \frac{\delta(0)}{96(4\pi)^{10/3}\kappa^{2/3}}
  \int_{M^{10}} d^{10}x \sqrt{g} \left( \chib^a \Gamma^{ABC} \chi^a \right)^2
\ee
$$
=
- \frac{1}{12\kappa^2} \int_{M^{11}} d^{11}x \sqrt{g}
  \left(G_{ABC11} 
        - \frac{\sqrt{2}}{16\pi} \left( \frac{\kappa}{4\pi} \right)^{2/3}
               \delta(x^{11}) \chib^a \Gamma_{ABC} \chi^a
  \right)^2
\,.
$$
After a careful calculation this leads to a vanishing variation 
$\delta\psi_A=0$. In our model (based on reduction and truncation) 
we can now compute these quantities explicitly. We assume gaugino 
condensation to occur at the $E_8$ boundary 
\be
\left<\chib^a \Gamma_{ijk} \chi^a \right> = \Lambda^3 \epsilon_{ijk}
\ee
where $\Lambda < M_{\rm GUT}$ and $\epsilon_{ijk}$ is the covariantly 
constant holomorphic 3--form. This leads to a nontrivial vev of 
$G_{11ABC}$ at this boundary and supersymmetry is 
broken\footnote
{One might speculate that a nontrivial vev of $D_A\eta$ might be operative 
here as in the case without gaugino condensation (see discussion after 
eq.\ (\ref{eq:dpsiM})). However, the special values of 
$H_{ijk} \propto \epsilon_{ijk}$ necessary to cancel the contribution of the 
gaugino condensate do not permit such a mechanism (see footnote 6 in ref.\ 
\cite{W}).}. 
At that boundary we obtain $F_\sS=0$ and $F_\sT \ne 0$ as expected from 
the fact that the component $\psi_{11}$ of the 11--dimensional gravitino 
plays the role of the goldstino.

In the effective $d=4$ theory we now have to average over the 11th 
dimension leading to
\be
\left< F_\sT \right> 
\approx
\frac{1}{2} \sT \frac{\int dx^{11} \delta\psi_{11}}{\int dx^{11}}
\ee
as the source of SUSY breakdown. This will then allow us to compute the 
size of supersymmetry breakdown on the observable $E_6$ side. Gravitational 
interactions play the role of messengers that communicate between the two 
boundaries. This effect can be seen from (\ref{eq:GN}): large $R_{11}$ 
corresponds to large $M_P$ and $\left< F_\sT \right>$ gives the effective 
size of SUSY breaking on the $E_6$ side ($R_{11} \rightarrow\infty$ implies 
$M_P \rightarrow\infty$). The gravitino mass is given by 
\be
m_{3/2} 
= {\left< F_\sT \right> \over \sT + \bar \sT} 
\approx {\Lambda^3 \over M_P^2}
\ee
(similar to (\ref{eq:m32}) in the weakly coupled case) and we expect this 
to represent the scale of soft supersymmetry breaking parameters in the 
observable sector. These soft masses are determined  by the coupling of 
the corresponding fields to the goldstino multiplet. As we have seen before, 
we cannot compute the scalar masses reliably in our approximation: 
$m_0=0$ because of the no--scale structure that appears as an artifact 
of our approximation. Fields of different modular weight will receive 
contribution to $m_0$ of order $m_{3/2}$. For the mass of a field 
$C$ we have  
\cite{KL,BIM}
\be
m_0^2 = m_{3/2}^2 - F^i {\bar F}^{\bar j} 
\frac{Z_{i \bar j} - Z_i Z^{-1} {\bar Z_{\bar j}}}{Z}
\ee
where $i,j = \sS, \sT$ and $Z$ is the moduli dependent coefficient of 
$C \bar C$ term appearing in the K\"ahler potential. Scalars of 
modular weight $-1$ will become massive through radiative corrections. 
This then leads to the expectation that $m_{3/2}$ should be in the 
TeV--region and $\Lambda\approx 10^{13}$ GeV\ \footnote
{In realistic models $E_8$ is broken and $\Lambda$ is adjusted by model 
building.}. 
So far this is all similar to the weakly coupled case.

An important difference appears, however, when we turn to the discussion 
of observable sector gaugino masses (\ref{eq:m12}). In the weakly coupled 
case they were zero at tree level and appeared only because of the 
radiative corrections at one loop (\ref{eq:f6f8}). As a result of this small 
correction, gaugino masses were expected to be much smaller than $m_{3/2}$. 
In the strongly coupled case the analog of (\ref{eq:m12}) is still valid 
\be
m_{1/2} = \frac{ {\partial f_6 \over \partial \sS} F_\sS 
                +{\partial f_6 \over \partial \sT} F_\sT }
               {2 {\rm Re} f_6}                          
\ee
and the 1--loop effect is encoded in the variation of the $\sigma$ and 
$\gamma$ fields from one boundary to the other. Here, however, the loop 
corrections are sizable compared to the classical result because of the 
fact that $R_{11}$ is close to its critical value. As a result we expect 
observable gaugino masses of the order of the gravitino mass. The problem 
of the small gaugino masses does therefore not occur in this situation. 
Independent of the question whether $F_\sS$ or $F_\sT$ are the dominant 
sources of supersymmetry breakdown, the gauginos will be heavy of the
order of the gravitino mass. The exact relation between the soft breaking 
parameters $m_0$ and $m_{1/2}$ will be a question of model building. If 
in some models $m_0 \ll m_{1/2}$ this might give a solution to the flavor 
problem. The no--scale structure found above might be a reason for such 
a suppression of $m_0$. As we have discussed above this structure, however, 
is an artifact of our simplified approximation and does not survive in 
perturbation theory. At best it could be kept exact (but only for the 
fields with modular weight $-1$) in the $R_{11} \rightarrow \infty$ limit. 
The upper bound on $R_{11}$ precludes such a situation. With observable 
gaugino masses of order $m_{3/2}$ we also see that $m_{3/2}$ cannot be 
arbitrarily large and should stay in the TeV -- range.

In recent months several other groups have studied similar questions in 
detail 
\cite{AQ,LLN,DG}. 
The discussion was explicitly done at the classical level. Some 
conclusions different from ours (concerning large values of 
$m_{3/2}$ and/or $R_{11}$) can only be obtained in that 
approximation. The one loop 
corrections, however, require $R_{11} < R_{\rm critical}$ as well as 
$m_{3/2}$ in the TeV -- range.

The picture of supersymmetry breakdown in the M--theoretic limit therefore 
seems very promising. It is very similar to the weakly coupled case, 
but avoids the problem of the small gaugino masses.

\vskip10mm
\noindent
{\Large \bf Acknowledgments}
\vskip5mm

This work was supported by 
the European Commission programs ERBFMRX--CT96--0045 and CT96--0090  
and by a grant from Deutsche Forschungsgemeinschaft SFB--375--95. 
The work of M.O. was partially supported by 
the Polish State Committe for Scientific Research grant 2 P03B 040 12. 
The work of M.Y. was partially supported by 
the Grant--in--Aid for Scientific Research from the Ministry of 
Education, Science and Culture of Japan No.\ 09640333.


\end{document}